\documentclass{ptephy_v1}

\usepackage{hyperref}
\hypersetup{
    unicode=false,          
    pdftoolbar=true,        
    pdfmenubar=true,        
    pdffitwindow=false,     
    pdfstartview={FitH},    
    pdfnewwindow=true,      
    colorlinks=true,       
    linkcolor=red,          
    citecolor=red,        
    filecolor=cyan,         
    urlcolor=magenta        
}

\usepackage[all]{xypic}
\usepackage{float}
\usepackage{tablefootnote}
\usepackage{textcomp}

\newcommand{\tmatrixfree}{{\it $t-$matrix-free~}}
\newcommand{\tmatrixdependent}{{\it $t-$matrix-dependent~}}

\newcommand{\tmatrixLS}{{\it $t-$matrix-LS~}}
\newcommand{\tmatrixFR}{{\it $t-$matrix-FR~}}

\newcommand{\bp}{{\bf p}}
\newcommand{\bpp}{\bp'}
\newcommand{\bpz}{\bp''}
\newcommand{\bk}{{\bf k}}
\newcommand{\bkp}{{\bf k}'}

\newcommand{\bq}{{\bf q}}

\newcommand{\half}{\frac{1}{2}}

\newcommand{\kett}{\rangle}
\newcommand{\bpi}{\boldsymbol \pi}
\newcommand{\btpi}{\tilde{\boldsymbol \pi}}

\begin{document}

\title{Relativistic Faddeev 3D Equations for Three-Body Bound States Without Two-Body $t-$Matrices}


\author[1]{M.~Mohammadzadeh}
\author[1]{M.~Radin}
\author[2,3]{M.~R.~Hadizadeh}
\affil[1]{Department of Physics, K. N. Toosi University of Technology, P.O.Box 163151618, Tehran, Iran,}
\affil[2]{College of Engineering, Science, Technology and Agriculture, Central State University, Wilberforce,
OH 45384, USA,}
\affil[3]{Department of Physics and Astronomy, Ohio University, Athens, OH 45701, USA,
\email{mhadizadeh@centralstate.edu}
}



\begin{abstract}%
This paper explores a novel revision of the Faddeev equation for three-body (3B) bound states, as initially proposed in Ref. \cite{golak2013three}. This innovative approach, referred to as \tmatrixfree in this paper, directly incorporates two-body (2B) interactions and completely avoids the 2B transition matrices. 
We extend this formalism to relativistic 3B bound states using a three-dimensional (3D) approach without using partial wave decomposition. To validate the proposed formulation, we perform a numerical study using spin-independent Malfliet-Tjon and Yamaguchi interactions. Our results demonstrate that the relativistic \tmatrixfree Faddeev equation, which directly implements boosted interactions, accurately reproduces the 3B mass eigenvalues obtained from the conventional form of Faddeev equation, referred to as \tmatrixdependent in this paper, with boosted 2B $t-$matrices. Moreover, the proposed formulation provides a simpler alternative to the standard approach, avoiding the computational complexity of calculating boosted 2B $t-$matrices and leading to significant computational time savings.
\end{abstract}

\subjectindex{xxxx, xxx}

\maketitle

\section{Introduction}
\label{sec:introduction}
Studying few-body quantum mechanical systems using the Faddeev-Yakubovsky (FY) scheme requires rigorous mathematical formalism and advanced computational techniques. The FY equations provide a non-perturbative formulation of the Schr\"odinger equation for three- and four-body problems, offering a reliable means to obtain converged and unique solutions. The solution of FY equations for bound and scattering systems across different sectors of physics, including atomic, nuclear, and particle physics, has been the subject of numerous studies in both momentum \cite{golak2013three,glockle1993alpha, stadler1997relativistic, elster1999three, nogga2002alpha, platter2004four, Keister2006, hadizadeh2007four, bayegan2008three, deltuva2010efimov, eichmann2010nucleon, witala2011three, deltuva2011universality, hadizadeh2011scaling, hadizadeh2012binding, hadizadeh2014relativistic, filikhin2018trions, mohseni2021three, etminan2022three, mohseni2022trion} and configuration \cite{payne1980configuration, fedorov1993efimov, bernabeu1996new, roudnev2000investigation, filikhin2002faddeev, thompson2004face, carbonell2011some, lazauskas2020description} spaces.
The inputs for the FY equations are 2B transition matrices $t(\epsilon)$, which are obtained by solving the inhomogeneous Lippmann-Schwinger (LS) equation for positive (scattering) or negative (bound states) 2B subsystem energies $\epsilon$.
However, solving the LS equation for positive energies can be numerically challenging due to singularities. When dealing with negative energies, it necessitates solving the LS equation for 2B subsystem energies determined by the Jacobi momenta of the third and fourth particles.
In Ref. \cite{golak2013three}, the authors present a novel formulation of Faddeev equations for a nonrelativistic 3B bound state that directly employs 2B interactions and entirely avoids using 2B $t-$matrices. In our study, we expand upon this scheme to establish the relativistic Faddeev integral equation that utilizes boosted 2B interactions as input within a 3D momentum representation, eliminating the need for boosted 2B $t-$matrices. This proposed formulation simplifies the formalism and the associated coding for implementing boosted 2B $t-$matrices, resulting in less demand on computational resources and a reduction in CPU processing time.
We verify the effectiveness of the \tmatrixfree formalism by computing relativistic 3B binding energies using the boosted potentials obtained from Malfliet-Tjon and Yamaguchi 2B potentials. Our numerical results indicate that the \tmatrixfree formulation of the relativistic Faddeev integral equation, using input boosted 2B interactions, aligns excellently with the results obtained from the \tmatrixdependent relativistic Faddeev integral equations that use input boosted 2B $t-$matrices.
It should be emphasized that the direct use of interactions in Faddeev equations is reliable for 3B bound states. However, for scattering calculations, it is necessary to rely on a method that involves 2B $t-$matrices.

In Section \ref{sec.Faddeev}, we provide a brief overview of the nonrelativistic and relativistic Faddeev equations, both with and without the utilization of 2B $t$-matrices. In Section \ref{sec.results}, we present our numerical results for nonrelativistic and relativistic 3B binding energies obtained using boosted 2B potentials constructed from Malfleit-Tjon and Yamaguchi potentials. To assess the accuracy of our calculations, we calculate the expectation value of the 3B Hamiltonian and compare it with the corresponding energy eigenvalues. The efficiency of the \tmatrixfree method is evaluated through an analysis of computational CPU time, alongside a comparison with two other \tmatrixLS and \tmatrixFR methods.
Finally, in Section \ref{sec.conclusion}, we provide a summary of our work.
%
\section{Faddeev equations for 3B bound states without 2B $t-$matrices}
\label{sec.Faddeev}
In the following section, we briefly introduce the \tmatrixfree formulation of the Faddeev equations proposed in Ref. \cite{golak2013three}. This formulation is designed explicitly for nonrelativistic 3B bound states and introduces the direct utilization of 2B potentials, thereby eliminating the requirement of 2B $t$-matrices. Additionally, we present an extension of this formulation to relativistic 3B bound states.

\subsection{Nonrelativistic 3B bound states}
The nonrelativistic bound state of three identical particles, with mass $m$, interacting with pairwise interactions $v_{nr}$ is described by the Faddeev equation as follows \cite{faddeev1965mathematical}
\begin{equation}
\label{Eq:Faddeev}
\psi_t=G_0tP\psi_t,
\end{equation}
where $\psi_t$ is the Faddeev component of the 3B wave function with input 2B $t-$matrices defined by the LS equation $t = v_{nr} + v_{nr} G_0 t$. $G_0=(E-H_0)^{-1}$ is 3B free propagator and $P=P_{12}P_{23}+P_{13}P_{23}$ is the permutation operator.
In the \tmatrixfree formulation proposed in Ref. \cite{golak2013three}, one can rewrite the Faddeev equation \eqref{Eq:Faddeev} as
\begin{equation}
\label{Eq:Faddeev_new}
\psi_v = G_0 v_{nr} (1+P)\psi_v,
\end{equation}
where 2B interaction $v_{nr}$ is being directly utilized as input to the Faddeev equation, and consequently, there is no need for the 2B $t$-matrices. The representation of \tmatrixfree form of the Faddeev equation \eqref{Eq:Faddeev_new} in momentum space leads to the following 3D integral equation
\begin{equation}
\label{Eq:Faddeev_new_integral}
 \psi_v (\bp,\bq)  =  \frac{1}{E-\frac{p^2}{m} -  \frac{3q^2}{4m} }  \biggl [ \int\,\, d^3 q'\,
 v_{nr}^{sym} (\bp,\btpi) \,\psi_v (\bpi,\bq') + \frac{1}{2} \int \ d^3 p'\, v_{nr}^{sym} (\bp,\bp')\,\psi_v (\bp',\bq) \biggl ] ,
\end{equation}
with the symmetrized 2B interaction $v_{nr}^{sym} (\bp,\bp') = v_{nr} (\bp,\bp') + v_{nr} ( -\bp,\bp')$ and the shifted momentum arguments $\bpi=\bq+\frac{1}{2}\bq'$ and $\btpi=\frac{1}{2}\bq+\bq'$.
By having the matrix elements of 2B interaction $v_{nr}(\bp,\bp')$, we first calculate the symmetrized interaction $v_{nr}^{sym} (\bp,\bp')$ and then solve the 3D integral equation \eqref{Eq:Faddeev_new_integral} by Lanczos iterative technique to calculate 3B binding energy and Faddeev component.
For comparison, the \tmatrixdependent form of the Faddeev equation \eqref{Eq:Faddeev} projected in momentum space can be expressed as the following 3D integral equation \cite{elster1999three}
\begin{equation}
\label{Eq:Faddeev_Original_integral}
 \psi_t (\bp,\bq)  =  \frac{1}{E-\frac{p^2}{m} -  \frac{3q^2}{4m}}  \,\, \int d^3q'\,
 t^{sym} (\bp,\tilde{\bpi};\epsilon )\, \psi_t (\bpi,\bq') ,
\end{equation}
where the symmetrized 2B $t-$matrices are defined as $t^{sym} (\bp,\bp';\epsilon ) = t (\bp,\bp';\epsilon ) + t (- \bp,\bp';\epsilon )$.
A comparison of the \tmatrixfree and \tmatrixdependent forms of the Faddeev equations, {\it i.e.}, Eqs. \eqref{Eq:Faddeev_new_integral} and \eqref{Eq:Faddeev_Original_integral}, shows that while the \tmatrixfree form includes an additional term involving the integration of the symmetrized 2B interaction and Faddeev component without interpolations on momenta or angles, its numerical solution is straightforward and less expensive than that of the \tmatrixdependent form. The \tmatrixdependent form requires the solution of the LS equation to calculate the fully-off-shell (FOS) 2B $t-$matrices for all 2B subsystem energies $\epsilon = E - \frac{3q^2}{4m}$ in each search for the 3B binding energy and for all values of Jacobi momentum $q$.

\subsection{Relativistic 3B bound states}
In relativistic quantum mechanics, the 3B mass operator for three identical particles with mass $m$ and momentum $\bk_i$, interacting via pairwise interactions, is defined as
\begin{eqnarray} \label{eq.3N-Hamiltonian}
M=M_0+\sum_{i<j} v_{k_{ij}},
\end{eqnarray}
where $M_0$ represents the free mass operator and $v_{k_{ij}} \equiv v_k$ are 2B boosted interactions embedded in the 3B Hilbert space. Following Kamada and Gl\"ockle approach \cite{kamada2007realistic}, the 2B boosted interaction $v_k$ can be obtained from the nonrelativistic 2B interactions $v_{nr}$ by solving a nonlinear equation given as $4m v_{nr} = \left ( \omega_k(p) v_k + v_k \omega_k(p) + v_k^2 \right)$, where $\omega_k(p)=\bigl (\omega^2(p)+k^2 \bigr)^{\half}$ and $\omega(p)=2\sqrt{m^2+p^2}$. Here, $\bk=\bk_i+\bk_j$ represents the total momentum of the subsystem $(ij)$, and $\bp$ is the relative momentum in the 2B subsystem $(ij)$.

The \tmatrixfree form of the relativistic Faddeev equation for describing three identical particles bound states is similar to the nonrelativistic case, and it can be expressed as
\begin{equation} \label{eq.Faddeev-rel_new}
 \psi_v=G_0 \, v_k\, (1+P)\, \psi_v,
\end{equation}
where $G_0=(M_t-M_0)^{-1}$ is free propagator, and $M_t=E + 3m$ is the 3B mass eigenvalue.
The representation of the \tmatrixfree form of the Faddeev equation \eqref{eq.Faddeev-rel_new} in the relativistic basis states $| \bp \, \bk \kett $ leads to
\begin{eqnarray} \label{eq.Faddeev-integral-rel_new}
\psi_v (\bp , \bk ) &=&
\frac{1}{M_t-M_0(p,k)}  \biggl [   \int d^3 k' \, N(\bk,\bkp) \,v_k^{sym} \big(\bp,\btpi \big)
 \, \psi_v \bigl(\bpi,\bkp \bigr) \cr
 &+&  \frac{1}{2} \int \ d^3 p' v_k^{sym} (\bp,\bp') \psi_v (\bp',\bk) \biggl ] ,
\end{eqnarray}
with the free mass operator $M_0(p,k) = \bigl ( \omega^2(p) + k^2 \bigr )^{\half} + \Omega(k) $, $\Omega(k) = \left (m^2+k^2 \right )^{\half}$, and the symmetrized boosted 2B interactions $v_k^{sym} (\bp,\bp') = v_k (\bp,\bp') + v_k ( -\bp,\bp')$. The explicit form of the Jacobian function $N(\bk,\bk')$ and the shifted momenta $\bpi$ and $\btpi$ are given in Refs. \cite{hadizadeh2014relativistic, hadizadeh2020three}.
The matrix elements of boosted potential $v_k(\bp,\bpp)$ can be obtained from the nonrelativistic potential $v_{nr}(\bp,\bpp)$ by solving a quadratic integral equation using an iterative scheme \cite{hadizadeh2017calculation, hadizadeh2021nnrelativistic, mohseni2021three}
\begin{equation} \label{eq.v-boost}
v_k(\bp,\bpp) + \frac{1}{\omega_k(p) + \omega_k(p')} \int d^3 p'' \, v_k(\bp,\bpz) \ v_k(\bpz,\bpp) =
 \frac{4m \ v_{nr}(\bp,\bpp)}{\omega_k(p) + \omega_k(p')} .
\end{equation}
For comparison, the representation of the \tmatrixdependent form of the relativistic Faddeev equation, {\it i.e.} $\psi_t=G_0 \, t_k\, \psi_t$, in momentum space leads to the following 3D integral equation \cite{hadizadeh2014relativistic, hadizadeh2020three}
\begin{equation} \label{eq.Faddeev-integral-rel_original}
\psi_t (\bp , \bk ) =
\frac{1}{M_t-M_0(p,k)}\int d^3 k' \, N(\bk,\bkp) \,t_{k}^{sym} \big(\bp,\btpi;\epsilon \big)
 \, \psi_t \bigl(\bpi,\bkp \bigr),
\end{equation}
with symmetrized FOS boosted $t-$matrices $t_{k}^{sym}(\bp,\bpp;\epsilon) =t_{k}(\bp,\bpp;\epsilon) +t_{k}(\bp,-\bpp;\epsilon)$. 
There are two different methods for the calculation of boosted FOS 2B $t-$matrices. The first, referred to as \tmatrixLS in our paper, involves solving the relativistic LS equation with boosted interactions \cite{hadizadeh2020three}. The second method, named \tmatrixFR here, employs a two-step approach for calculating the boosted FOS 2B $t-$matrices \cite{hadizadeh2014relativistic}. Initially, the nonrelativistic right-half-shell (RHS) $t-$matrices are calculated via the solution of the LS equation and then an analytical process computes the boosted RHS $t-$matrices. Subsequently, by solving the first resolvent (FR) equation, the boosted FOS 2B $t-$matrices are derived from the RHS boosted $t-$matrices.
In this study, we utilize the \tmatrixLS as one of the \tmatrixdependent methods to assess the accuracy of our binding energy results obtained from the \tmatrixfree form of the Faddeev equation. To this aim, the boosted FOS 2B $t-$matrices $t_{k}(\bp,\bpp;\epsilon)$ can be obtained from the boosted potential $v_k(\bp,\bpp)$ by solving the relativistic LS equation for 2B subsystem energies $\epsilon=M_t- \Omega(k)$ \cite{hadizadeh2017calculation}
\begin{equation} \label{eq.t-matrix-boost}
t_{k}(\bp,\bpp;\epsilon)=v_k(\bp,\bpp)+ \int d^3 p'' \, \frac{v_k(\bp,\bpz)}
{\epsilon - \bigl (\omega^2(p'')+ k^2 \bigr )^{\half}}\,t_{k}(\bpz,\bpp;\epsilon),
\end{equation}
Comparing the \tmatrixfree and \tmatrixdependent forms of relativistic Faddeev integral equations, {\it i.e.}, Eqs. \eqref{eq.Faddeev-integral-rel_new} and \eqref{eq.Faddeev-integral-rel_original}, it is clear that the \tmatrixfree form despite an additional term that integrates over the symmetrized 2B boosted interaction and Faddeev component, offers a more straightforward computational approach for coding and implementing. 
Conversely, the \tmatrixdependent form requires calculating the boosted FOS 2B $t$-matrices for all 2B subsystem energies $\epsilon = M_t - \Omega(k)$ at each step of the search process in the 3B mass eigenvalue $M_t$, for every value of the Jacobi momentum $k$.

\subsection{Summary of the formalism}
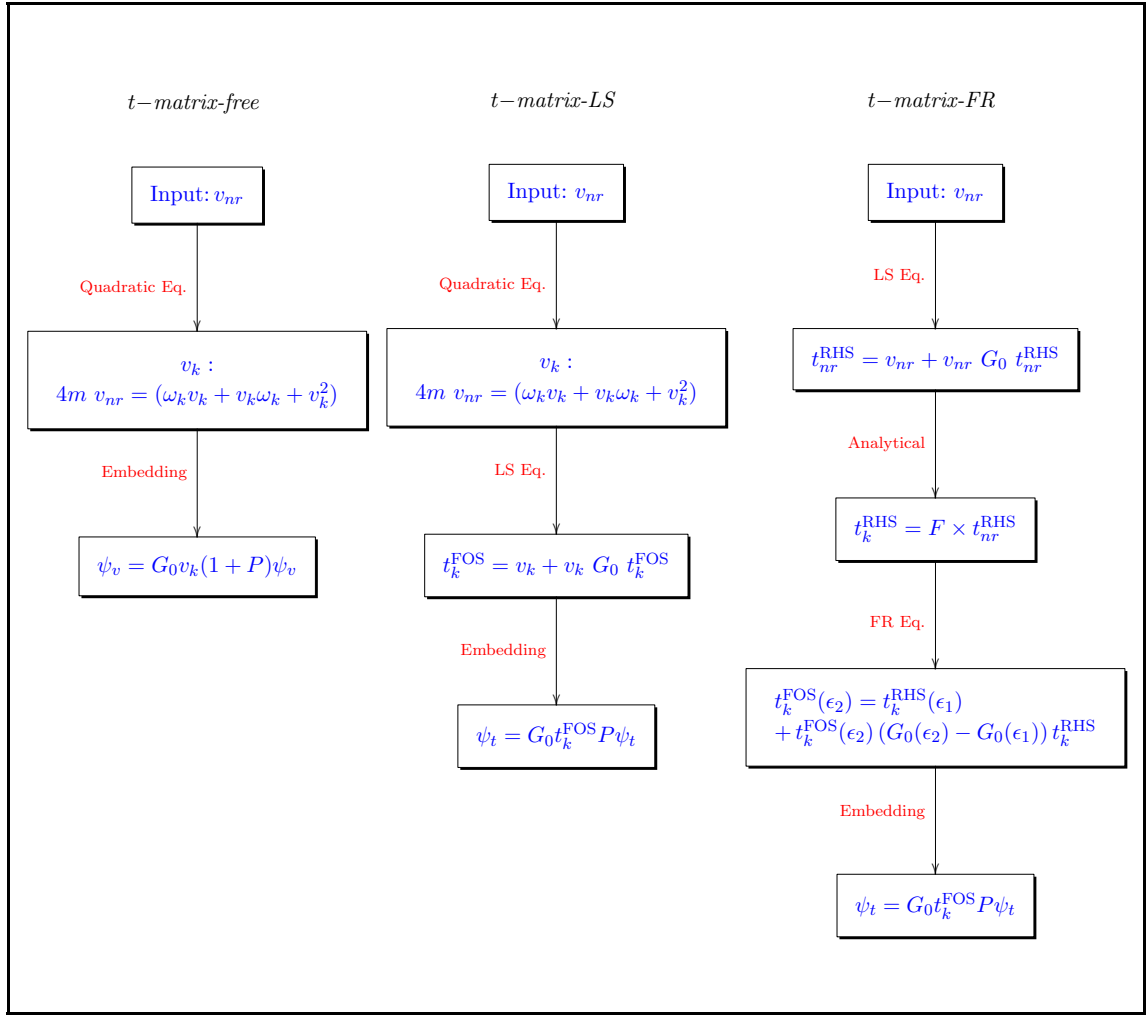
\begin{figure}[H]
  \begin{displaymath}
    \resizebox{1.\linewidth}{!}{
    \fboxrule=.5mm
  \boxed{
  \begin{array}{ccc}
  \xymatrix @-0.2pc{
    \\
     {\textbf{\tmatrixfree}}
   \\
    *+<18pt>[F-,]{{\color{blue}\text{Input:}\, v_{nr}}} 
   \ar[dd]_{{\color{red}\text{Quadratic Eq.}}} 
   \\ \\
   *+<18pt>[F-,]
   {
   {\color{blue}
  \begin{array}{c} 
 v_k: \\ 
 4m \ v_{nr} = (\omega_k v_k + v_k \omega_k + v_k^2)
    \end{array}
   }
  }
   \ar[dd]_{{\color{red}\text{Embedding}}} 
   \\ \\
   *+<18pt>[F-,]{{\color{blue}\psi_v = G_0v_k(1+P)\psi_v}}
   \\
  }  
  &
  \xymatrix @-0.2pc{
     \\
      {\textbf{\tmatrixLS}}
   \\
    *+<18pt>[F-,]{{\color{blue}\text{Input:}\,\,v_{nr}}} 
   \ar[dd]_{{\color{red}\text{Quadratic Eq.}}} 
   \\ \\
    *+<18pt>[F-,]
   {
   {\color{blue}
  \begin{array}{c} 
 v_k: \\ 
 4m \ v_{nr} = (\omega_k v_k + v_k \omega_k + v_k^2)
    \end{array}
   }
  } 
   \ar[dd]_{{\color{red}\text{LS Eq.}}} 
   \\ \\
   *+<18pt>[F-,]{{\color{blue} t^{\text{FOS}}_k = v_k + v_k \ G_0 \ t^{\text{FOS}}_k
   }}  \ar[dd]_{{\color{red}\text{Embedding}}}
   \\ \\
   *+<18pt>[F-,]{{\color{blue}\psi_t=G_0 t^{\text{FOS}}_k P\psi_t}}
    \\ \\
  } 
  & 
  \xymatrix @-0.2pc{
   \\
     {\textbf{\tmatrixFR}}
   \\
    *+<18pt>[F-,]{\color{blue} \text{Input:}\,\, v_{nr}} 
   \ar[dd]_{\color{red}\text{LS Eq.}} 
   \\ \\
   *+<18pt>[F-,]{\color{blue} t^{\text{RHS}}_{nr} = v_{nr} + v_{nr} \ G_0 \ t^{\text{RHS}}_{nr} } 
   \ar[dd]_{\color{red}\text{Analytical}} 
   \\\\
   *+<18pt>[F-,]{\color{blue} t^{\text{RHS}}_k = F \times t^{\text{RHS}}_{nr}}  
   \ar[dd]_{\color{red}\text{FR Eq.}}
   \\ \\
   *+<18pt>[F-,]
   {
   {\color{blue}
  \begin{array}{l} 
 t^{\text{FOS}}_k (\epsilon_2) = t^{\text{RHS}}_k (\epsilon_1) \\ 
 + \, t^{\text{FOS}}_k (\epsilon_2) \, (G_0(\epsilon_2) - G_0 (\epsilon_1) ) \, t^{\text{RHS}}_k
    \end{array}
   }
  } 
   \ar[dd]_{\color{red}\text{Embedding}}
    \\ \\
   *+<18pt>[F-,]{\color{blue}\psi_t=G_0 t^{\text{FOS}}_k P\psi_t} 
    \\ \\
  }
  \\
  \end{array}
  }
    } 
  \end{displaymath}
\caption{
Comparison of three different methods for solving relativistic Faddeev equations for 3B bound states. Left Panel: The \tmatrixfree method, initiating with a nonrelativistic potential $v_{nr}$, numerically computes the boosted potential $v_k$, by solving a quadratic equation, to be embedded in the relativistic Faddeev equation $\psi_v = G_0v_k(1+P)\psi_v$. Middle Panel: In the \tmatrixLS approach, an additional step calculates the boosted FOS $t$-matrices $t_k^{FOS}$ from $v_k$, via the LS equation, before their incorporation into the Faddeev equation $\psi_t=G_0 t^{\text{FOS}}_k P\psi_t$. Right Panel: The \tmatrixFR method first calculates the RHS nonrelativistic $t$-matrices $t^{RHS}_{nr}$ from $v_{nr}$ using the LS equation, followed by an analytical derivation of the boosted RHS $t$-matrices $t^{RHS}_{k}$ from $t^{RHS}_{nr}$ with the momentum-dependent function $F$. Subsequently, the FR equation is solved to obtain the FOS $t$-matrices $t_k^{FOS}$ for embedding into the relativistic Faddeev equation $\psi_t=G_0 t^{\text{FOS}}_k P\psi_t$.
}
 \label{flowchart}
 \end{figure}

Table \ref{table:Fadeev_summary} summarizes the explicit operator forms of the \tmatrixfree and \tmatrixdependent (either \tmatrixLS or \tmatrixFR) Faddeev equations for 3B bound states in both nonrelativistic and relativistic schemes. As shown in the table, the \tmatrixfree formulation has a more direct approach by utilizing the 2B nonrelativistic and boosted interactions. Conversely, the \tmatrixdependent requires additional steps for calculating the FOS $t-$matrices. Specifically, in the nonrelativistic case, the \tmatrixdependent form involves an extra step for calculating the FOS $t-$matrices by solving the LS equation, as can be done in the \tmatrixLS method. However, in the relativistic case, this approach expands to include either solving the LS equation (in the \tmatrixLS method) or addressing the FR equation (in the \tmatrixFR method) to calculate the boosted FOS $t-$matrices.
\begin{table}[hbt]
\caption{The summary of the \tmatrixfree and \tmatrixdependent forms of nonrelativistic (NR) and relativistic (R) Faddeev equations and their inputs for 3B bound states. As discussed in the text, the \tmatrixdependent form can be addressed using either the \tmatrixLS (LS) method or the \tmatrixFR (FR) method.}
\centering
\begin{tabular}{c|cccccc|}
\hline
& Faddeev equation & inputs && Faddeev equation & inputs \\
\hline
   &  \multicolumn{2}{c}{\tmatrixfree}  &&  \multicolumn{2}{c}{\tmatrixdependent (LS)}    \\  \cline{2-3}   \cline{5-6}
NR & $\psi_v = G_0 v_{nr} (1+P)\psi_v$ & $v_{nr}$ && $\psi_t=G_0tP\psi_t$ & $t(E-\frac{3q^2}{4m})$\\
\hline
   &  \multicolumn{2}{c}{\tmatrixfree}  &&  \multicolumn{2}{c}{\tmatrixdependent (LS \& FR)}    \\  \cline{2-3}   \cline{5-6}
R & $\psi_v=G_0 \, v_k\, (1+P)\, \psi_v$ & $v_k$ && $\psi_t=G_0 \, t_k\, \psi_t$ & $t_k (M_t -  \sqrt{m^2+k^2} )$\\
\hline
\end{tabular}
\label{table:Fadeev_summary}
\end{table}
One should note that for the relativistic 3B bound states, while the \tmatrixfree method utilizes directly the boosted potentials $v_k$, in the \tmatrixdependent approach, two different strategies can be employed to calculate the boosted FOS $t-$matrices $t_k (M_t -  \sqrt{m^2+k^2})$. 
The flowchart diagram in Fig. \ref{flowchart} illustrates three methods for solving the relativistic Faddeev equation, each represented in a separate panel: \tmatrixfree on the left, \tmatrixLS in the middle, and \tmatrixFR on the right. All three methods begin with the same input nonrelativistic potential $v_{nr}$. 

\begin{itemize}

\item In the \tmatrixfree method (left panel), the input nonrelativistic potential $v_{nr}$ undergoes numerical processing to yield the boosted potential $v_k$ by solving a quadratic equation, followed by embedding in the relativistic Faddeev equation $\psi_v = G_0 v_k(1+P) \psi_v$.

\item The \tmatrixLS method (middle panel) also starts with a numerical solution of a quadratic equation to obtain the boosted potential $v_k$ from the input nonrelativistic potential $v_{nr}$. Then, $v_k$ is utilized to solve the relativistic LS equation to calculate the boosted FOS $t-$matrices $t^{\text{FOS}}_k$, which is embedded in the Faddeev equation $\psi_t=G_0 t^{\text{FOS}}_k P \psi_t$.

\item Finally, in the \tmatrixFR method (right panel), the input $v_{nr}$ is utilized to solve the LS equation and calculate the nonrelativistic RHS $t-$matrices $t^{RHS}_{nr}$. Then the boosted RHS $t-$matrices $t^{RHS}_k$ are obtained analytically from $t^{RHS}_{nr}$ using an analytical momentum-dependent function $F$. This is followed by solving the FR equation to calculate the boosted FOS $t-$matrices $t^{\text{FOS}}_k$ and finally embedding them into the relativistic Faddeev equation $\psi_t=G_0t_kP\psi_t$. It should be noted that in this method there is no need to calculate the boosted potentials, however, the solution of the FR equation demands treating the moving singularities.

\end{itemize}

The flowchart diagrams clearly illustrate that the \tmatrixfree approach is more direct in its formalism and coding. In comparison, the \tmatrixLS requires an additional step to solve the LS equation for the boosted FOS $t$-matrices. Furthermore, the \tmatrixFR necessitates a two-step numerical process to calculate the nonrelativistic RHS and boosted FOS $t$-matrices. 
In the next section, we will study and compare the computational time required for performing each of these methods.

\section{Numerical Results}
\label{sec.results}
\subsection{3B binding energies and expectation values}

To assess the accuracy of the \tmatrixfree formulation of Faddeev equations, we conduct our numerical tests with two spin-independent 2B interactions: Malfliet-Tjon V (MT-V) \cite{malfliet1969solution} and Yamaguchi model IV (Yamaguchi-IV) \cite{yamaguchi1954two}.
The details of MT-V and Yamaguchi-IV potential parameters can be found in Ref. \cite{hadizadeh2007four}.
Refs. \cite{elster1999three, hadizadeh2014relativistic, hadizadeh2020three} provide a comprehensive description of the numerical methods used to solve the Faddeev integral equations in both the nonrelativistic and relativistic cases.
Table \ref{table:Nonrel_3B_energiy_convergence} displays our numerical results for the convergence of the nonrelativistic and relativistic 3B binding energies as a function of the number of mesh points for Jacobi momenta using the \tmatrixdependent and \tmatrixfree forms of the Faddeev equations. Specifically, we solve Eqs. \eqref{Eq:Faddeev_new_integral} and \eqref{Eq:Faddeev_Original_integral} for the nonrelativistic case and Eqs. \eqref{eq.Faddeev-integral-rel_new} and \eqref{eq.Faddeev-integral-rel_original} for the relativistic case.

\begin{table}[hbt]
\caption{The convergence of nonrelativistic ($nr$) and relativistic ($r$) 3B binding energies obtained from the \tmatrixfree ($E^{nr}_{v}$ and $E^{r}_{v}$) and \tmatrixLS ($E^{nr}_{t}$ and $E^{r}_{t}$) forms of Faddeev equations as a function of the number of mesh points for Jacobi momenta $N_{jac}$. The results are presented for MT-V (left panel) and Yamaguchi-IV (right panel) potentials. We use 40 mesh points for both the polar and azimuthal angles. All the energies are given in MeV.}
\label{table:Nonrel_3B_energiy_convergence}
\begin{center}
\begin{tabular}{c|cccc|ccccccccc}
\hline
 $N_{jac}$   &  $E^{nr}_{v}$ &  $E^{nr}_{t}$ &  $E^{r}_{v}$ &  $E^{r}_{t}$ & $E^{nr}_{v}$ &  $E^{nr}_{t}$ &  $E^{r}_{v}$ &  $E^{r}_{t}$  \\ \hline
&  \multicolumn{4}{c}{MT-V} &  \multicolumn{4}{c}{Yamaguchi-IV}  \\  \cline{2-5} \cline{6-9}
20  &    -7.9557   &  -7.9562  &    -7.8049  &  -7.8050   &  -8.4983 &  -8.4984  &   -8.2955 &  -8.2956 \\
30  &   -7.7966  &  -7.7971  &   -7.6442   &  -7.6442   & -8.5057 &  -8.5058  &   -8.3023 &  -8.3023 \\
40  &   -7.7468  &  -7.7475  &  -7.5952  &  -7.5952   &  -8.5078 &  -8.5078  &   -8.3041 &  -8.3042  \\
50  &    -7.7465   &  -7.7473  &  -7.5950   &  -7.5951   &  -8.5084 &  -8.5085  &   -8.3047 &  -8.3048  \\
60  &    -7.7392   &  -7.7399  &   -7.5877   &  -7.5877   & -8.5086 &  -8.5087  &   -8.3049 &  -8.3050 \\
70  &   -7.7391   &  -7.7398   &   -7.5877   &  -7.5877   &  -8.5087 &  -8.5088  &   -8.3050 &  -8.3051 \\
80  &    -7.7372  &  -7.7379   &  -7.5858  &   -7.5858   &  -8.5088 &  -8.5089  &   -8.3051 &  -8.3051 \\
90  &    -7.7364   &  -7.7371  &    -7.5850  &   -7.5851   &  -8.5088 &  -8.5089  &   -8.3051 &  -8.3052  \\
100  &    -7.7361  &  -7.7368   &  -7.5847    &  -7.5848   &  -8.5088 &  -8.5089  &   -8.3051 &  -8.3052  \\
110  &    -7.7360  &  -7.7368$^1$   &   -7.5847   &  -7.5848   &  -8.5088 &  -8.5089  &   -8.3051 &  -8.3052 \\
\hline
\end{tabular}
\end{center}
\footnotesize{$^1$ The converged nonrelativistic binding energy  $-7.7368$ MeV obtained using the \tmatrixLS method shows excellent agreement with the binding energy $-7.7365$ MeV reported in Ref. \cite{elster1999three}. However, it differs slightly from the $-7.7382$ MeV reported in Ref. \cite{hadizadeh2014relativistic}, which was obtained using a lower number of mesh points, specifically 100 mesh points for Jacobi momentum $p$ and $60$ mesh points for Jacobi momentum $k$.} 
\end{table}
As shown in Table \ref{table:Nonrel_3B_energiy_convergence}, the nonrelativistic and relativistic 3B binding energies obtained from the \tmatrixfree form of the Faddeev equations using the MT-V and Yamaguchi-IV potentials are in excellent agreement with the corresponding energies obtained from the \tmatrixdependent form of Faddeev equations. 
It is evident that in all cases, the relative percentage difference in 3B binding energies obtained from the \tmatrixfree and \tmatrixLS methods is less than $0.01$\%. Specifically, the energy difference is about $0.1$ keV for each case, except in the nonrelativistic MT-V scenario, where the difference is about $0.8$ keV. When solving Faddeev integral equations, each iteration step requires two-dimensional interpolations on a shifted momentum and an angle, either on the potential or $t-$matrices. In the nonrelativistic \tmatrixfree approach, this interpolation is bypassed due to the analytical form of the potential, whereas it must be performed numerically for nonrelativistic $t-$matrices, leading to a higher difference in binding energies. For relativistic cases, two-dimensional interpolations are necessary and performed numerically in both the \tmatrixfree and \tmatrixLS methods, resulting in a similar level of accuracy or error. 
It should be noted that the 3B binding energies obtained from the Yamaguchi-IV potential achieve rapid convergence using a relatively small number of mesh points for Jacobi momenta at around $N_{jac} = 70$. This quick convergence when using the Yamaguchi potential, in comparison to the Malfliet-Tjon potential, is due to its $s-$wave separable structure, with no angular dependency.
\begin{table}[hbt]
\caption{The upper panel displays the expectation values of the nonrelativistic 3B Hamiltonian, $\langle H \rangle$, obtained from the 3B free Hamiltonian, $\langle H_0 \rangle$, and the 2B interactions, $\langle V \rangle$. The 3B Hamiltonian expectation values are compared to the nonrelativistic 3B binding energy, $E^{nr}$, calculated for MT-V and Yamaguchi-IV potentials using both the \tmatrixfree and \tmatrixLS forms of the Faddeev equations. The lower panel shows the corresponding results for the relativistic case. Specifically, we display the expectation values of the 3B mass operator, $\langle M \rangle$, obtained from the free mass operator, $\langle M_0 \rangle$, and the boosted 2B interactions, $\langle V_k \rangle$. The 3B mass operator expectation values are compared to the relativistic 3B binding energy, $E^r$. All numbers are given in units of MeV.}
\centering
\begin{center}
\begin{tabular}{c|ccccccccccccc}
\hline
Faddeev scheme  &  $\langle H_0 \rangle$ &   $\langle V \rangle$ &  $\langle H \rangle$ &  $E^{nr}$ &  $ \left | \frac{\langle H \rangle - E^{nr}}{E^{nr}} \right | \times 100 \%$ \\ \hline
 &  \multicolumn{5}{c}{MT-V} \\
   \cline{2-6}
\tmatrixfree &  29.7659 &  -37.5037  &  -7.7377 &   -7.7360 &  0.0220  \\
\tmatrixLS &  29.7704 &  -37.5080 &  -7.7376 &  -7.7368 &  0.0103  \\
 \hline
  &  \multicolumn{5}{c}{Yamaguchi-IV} \\
    \cline{2-6}
\tmatrixfree  &  33.8678 &  -42.3786 &  -8.5107 &  -8.5088 &  0.0223 \\
\tmatrixLS &  33.8682 &  -42.3790  &  -8.5107 &  -8.5089 &  0.0211 \\
\hline \hline
 &  $\langle M_0 \rangle - 3m$ &   $\langle V_k \rangle$ &  $\langle M \rangle -3m $ &  $E^r$ &  $ \left | \frac{\langle M \rangle - 3m - E^r}{E^r} \right | \times 100 \%$ \\
 \hline
  &  \multicolumn{5}{c}{MT-V} \\
   \cline{2-6}
 \tmatrixfree  &  28.3824 &  -35.9656 &  -7.5832 &  -7.5847 &  0.0198 \\
 \tmatrixLS  &  28.3830 &  -35.9661 &  -7.5832 &  -7.5848 &  0.0211 \\
 \hline
 &   \multicolumn{5}{c}{Yamaguchi-IV} \\
    \cline{2-6}
 \tmatrixfree &  32.6517 &  -40.9572 &  -8.3055 &  -8.3051 &  0.0048 \\
 \tmatrixLS &  32.6518 &  -40.9574 &   -8.3056 &  -8.3052 &  0.0048 \\
 \hline
\end{tabular}
\end{center}
\label{table:3B_expectation}
\end{table}
To complement our assessment of the validity of the \tmatrixfree formulation and the quality of the 3B wave function, we also calculate the expectation values of the nonrelativistic and relativistic 3B Hamiltonian. These expectation values are presented in Table \ref{table:3B_expectation} for MT-V and Yamaguchi-IV potentials. Our comparison between the expectation value of the Hamiltonian and the calculated energy eigenvalues for both nonrelativistic and relativistic bound states shows that the \tmatrixfree form of the Faddeev equations provides the same level of accuracy as the \tmatrixdependent form. The relative percentage difference between the 3B binding energies and the corresponding expectation value of the 3B Hamiltonian, for both nonrelativistic and relativistic scenarios, is equal to or less than $0.022 \%$. This indicates that the \tmatrixfree forms of the Faddeev equations are able to perfectly reproduce the results of the \tmatrixdependent forms while avoiding the complexity associated with the calculation of 2B $t-$matrices.
The ability to avoid the calculation of 2B $t-$matrices significantly simplifies the computational procedure for calculating 3B binding energies. As a result, the \tmatrixfree forms of the Faddeev equations provide an efficient approach to solving the 3B bound state problems in both nonrelativistic and relativistic cases without sacrificing accuracy.

\subsection{CPU usage comparison: evaluating \tmatrixfree and \tmatrixdependent methods}

\begin{figure}[hbt]
\centering
\includegraphics[width=0.49\textwidth]{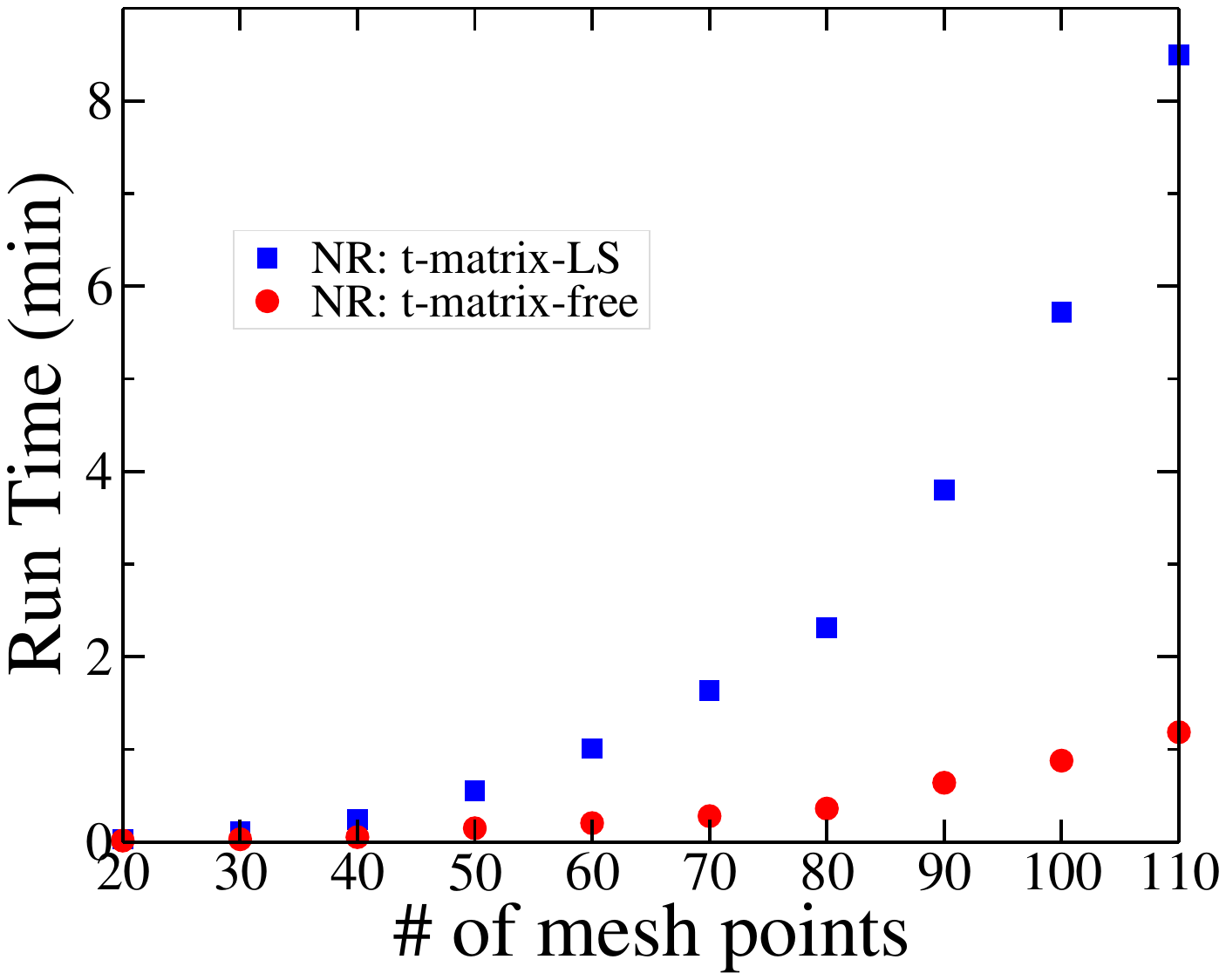}
\includegraphics[width=0.49\textwidth]{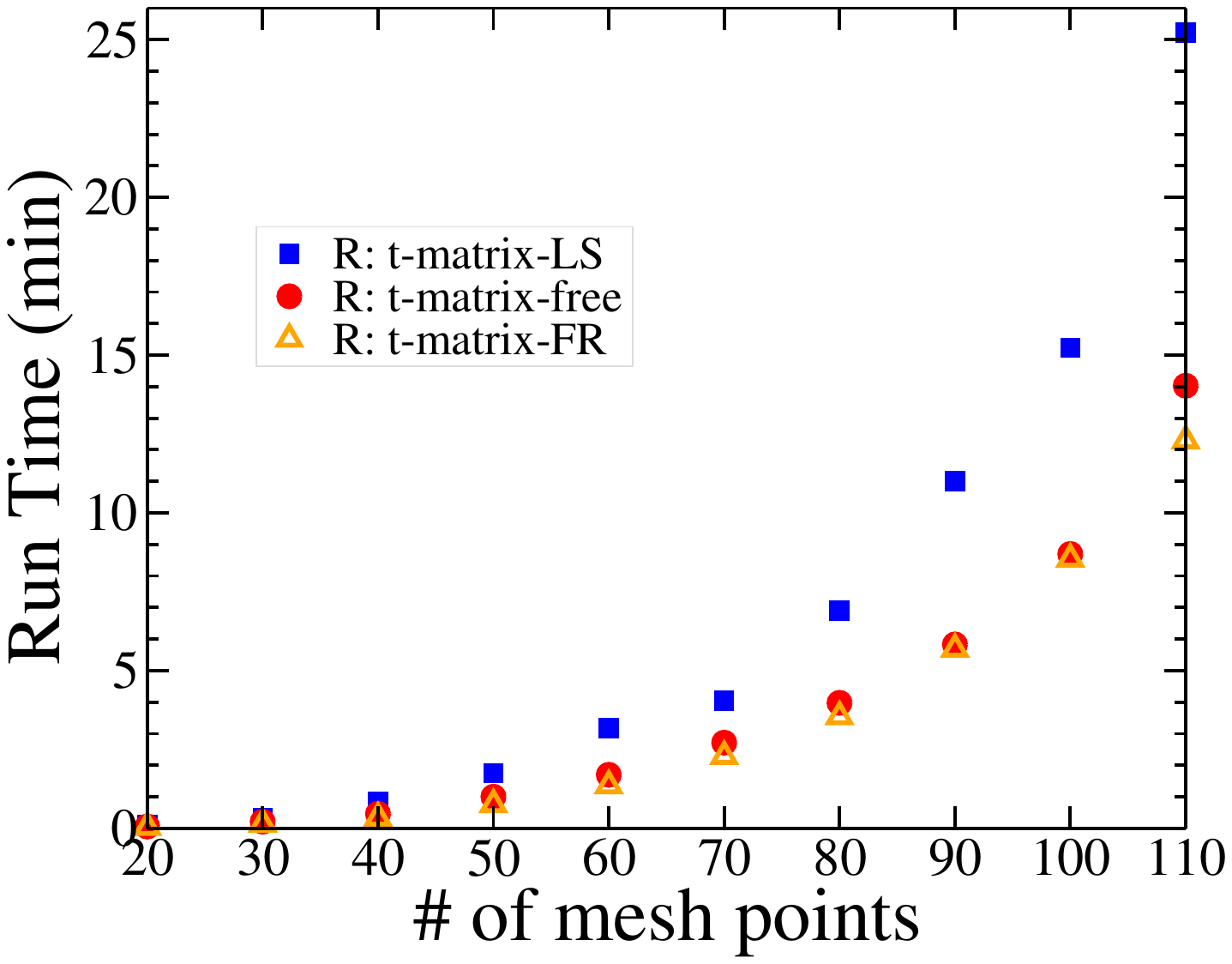}
\caption{Comparison of computational time per iteration in the energy search for 3B bound state calculations: The left panel represents the nonrelativistic (NR) scenario, detailing CPU time against the number of mesh points for the magnitude of Jacobi momenta for the \tmatrixLS (squares) and \tmatrixfree (circles) methods. The right panel illustrates the relativistic (R) case, displaying the CPU time for both the \tmatrixLS (squares) and \tmatrixFR (triangles) methods, compared with the \tmatrixfree (circles) method, providing a direct comparison of the computational time demands of each method.}
\label{runtime}
\end{figure}

While the proposed \tmatrixfree form of the Faddeev integral equations accurately reproduces the results of the \tmatrixdependent form, our numerical analysis demonstrates that the \tmatrixfree form significantly reduces the computational time required for each energy search. In Fig. \ref{runtime}, we compare the computational time per iteration in each search for energy for 3B bound state calculations. 
The calculations are conducted using a workstation equipped with Intel$^{\text{\textregistered}}$ Xeon$^{\text{\textregistered}}$ Processors E5-2680 v4, which feature a 35M Cache and a clock speed of 2.40 GHz.
In the left panel, which illustrates the nonrelativistic (NR) case, we observe that as the number of mesh points for the magnitude of Jacobi momenta increases, there is a corresponding increase in the CPU time for both the \tmatrixLS (squares) and the \tmatrixfree method (circles). However, the results demonstrate that at higher mesh points, the \tmatrixfree method considerably reduces CPU time. For instance, in the case of 110 mesh points, the computational time for the \tmatrixfree method is about 7 times lower than that required for the \tmatrixLS method. This significant decrease in time highlights the effectiveness of the \tmatrixfree method, especially when dealing with larger numbers of mesh points.
The right panel shows the results for the relativistic (R) scenario. Here, both \tmatrixdependent methods, {\it i.e.}, \tmatrixLS, denoted by squares, and the \tmatrixFR, denoted by triangles, are compared against the \tmatrixfree method (circles). 
It is noteworthy that the \tmatrixFR and \tmatrixfree methods display a similar level of computational demand, as evidenced by their nearly identical run times at comparable mesh points. In contrast, the \tmatrixLS method shows a significant increase in computational time, especially at higher mesh points. For instance, at 110 mesh points, the \tmatrixLS requires approximately 1.8 times more computational time compared to the \tmatrixfree method. Meanwhile, the \tmatrixfree method's computational time is only about 1.1 times higher than that of the \tmatrixFR.
In both panels, the \tmatrixfree method consistently shows the least computational time with respect to \tmatrixLS, indicating its potential advantage in terms of efficiency for practical computations. 
Such a comparative analysis provides valuable insights on how the \tmatrixfree method can achieve both high accuracy and computational efficiency, offering valuable insights for refining numerical methods by computational physicists who are exploring quantum 3B bound states.

\section{Summary}
\label{sec.conclusion}
In this study, we extend upon the \tmatrixfree formulation of the Faddeev equation proposed in Ref. \cite{golak2013three} to enable a relativistic description of 3B bound states. Our approach directly incorporates 2B boosted interactions and eliminates the need for the FOS 2B $t$-matrices. This simplification enhances computational efficiency and provides an advantage over the \tmatrixdependent formulation, which necessitates solving the LS equation for 2B subsystem energies, depending on the second Jacobi momentum, at each step of the 3B binding energy search.
To assess the effectiveness of our approach, we perform numerical tests on both nonrelativistic and relativistic 3B bound states. Using two spin-independent Malfliet-Tjon and Yamaguchi potentials, we employ a 3D scheme without partial wave decomposition. Our results, summarized in Table \ref{table:3B_expectation}, demonstrate that the \tmatrixfree formulation of the Faddeev equations yields comparable accuracy to the \tmatrixdependent formulation while avoiding the complexities associated with coding and implementing 2B $t$-matrices. 
Moreover, as shown in Figs. \ref{flowchart} and \ref{runtime}, the \tmatrixfree method presents more notable advantages, particularly in terms of simplicity of formalism and coding, as well as in reducing computational time. This method eliminates the need to calculate 2B $t-$matrices for all 2B subsystem energies, thereby significantly simplifying the formalism and calculations involved. 
These findings have notable implications for achieving efficient and accurate numerical solutions to relativistic 3B problems.
Furthermore, we emphasize the potential extension of this \tmatrixfree formulation to four-body (4B) bound states. Our preliminary results for 4B binding energies, obtained from the \tmatrixfree version of the coupled Yakubovsky equations using $s-$wave nonseparable (MT-V) and separable (Yamaguchi and Gaussian) interactions, show excellent agreement with the binding energies obtained from the \tmatrixLS form \cite{mohammadzadeh11four}. Efforts to extend these calculations to more general interactions are currently underway.

\section*{Acknowledgment}
We thank Kamyar Mohseni for the helpful discussions.
M.R. acknowledges the financial support from Iran National Science Foundation (INSF), under Grant No. 4003662.
The work of M.R.H. was supported by the National Science Foundation under Grant No. NSF-PHY-2000029 with Central State University.



\bibliographystyle{ptephy}
\bibliography{References}
%

\end{document}